\documentclass[conference]{IEEEtran}
\IEEEoverridecommandlockouts
\usepackage{cite}
\usepackage{amsmath,amssymb,amsfonts}
\usepackage{algorithmic}
\usepackage{graphicx}
\usepackage{textcomp}
\usepackage{xcolor}
\usepackage{hyperref}
\usepackage{listings}
\usepackage{float}
\usepackage{algorithmic}
\usepackage[linesnumbered,ruled,vlined]{algorithm2e}
\usepackage{tikz}
\usepackage{pgfplots}
\usepackage{balance}
\pgfplotsset{compat=1.18}

\def\BibTeX{{\rm B\kern-.05em{\sc i\kern-.025em b}\kern-.08em
    T\kern-.1667em\lower.7ex\hbox{E}\kern-.125emX}}

\newcommand\YAMLcolonstyle{\color{red}\mdseries}
\newcommand\YAMLkeystyle{\color{black}\ttfamily\footnotesize}
\newcommand\YAMLvaluestyle{\color{blue}\mdseries}

\makeatletter

\newcommand\language@yaml{yaml}

\expandafter\expandafter\expandafter\lstdefinelanguage
\expandafter{\language@yaml}
{
  keywords={true1,false1,null1,y1,n1},
  keywordstyle=\color{darkgray}\bfseries,
  basicstyle=\YAMLkeystyle,                                 
  sensitive=false,
  comment=[l]{\#},
  morecomment=[s]{/*}{*/},
  commentstyle=\color{purple}\ttfamily,
  stringstyle=\YAMLvaluestyle\ttfamily,
  moredelim=[l][\color{orange}]{\&},
  moredelim=[l][\color{magenta}]{*},
  moredelim=**[il][\YAMLcolonstyle{:}\YAMLvaluestyle]{:},   
  morestring=[b]',
  morestring=[b]",
  literate =    {---}{{\ProcessThreeDashes}}3
                {>}{{\textcolor{red}\textgreater}}1
                {|}{{\textcolor{red}\textbar}}1
                {\ -\ }{{\mdseries\ -\ }}3,
  xleftmargin=2em,
}

\lst@AddToHook{EveryLine}{\ifx\lst@language\language@yaml\YAMLkeystyle\fi}
\makeatother

\begin{document}

\title{HPCAdvisor: A Tool for Assisting Users in Selecting HPC Resources in the Cloud}

\author{\IEEEauthorblockN{Marco A. S. Netto}
\IEEEauthorblockA{\textit{Microsoft} \\
Redmond, USA\\marconetto@microsoft.com}

}

\maketitle

\begin{abstract}

Cloud platforms are increasingly being used to run HPC workloads. Major cloud
providers offer a wide variety of virtual machine (VM) types, enabling
users to find the optimal balance between performance and cost. However,
this extensive selection of VM types can also present challenges, as users
must decide not only which VM types to use but also how many nodes are
required for a given workload. Although benchmarking data is available for
well-known applications from major cloud providers, the choice of resources
is also influenced by the specifics of the user's application input. This
paper presents the vision and current implementation of HPCAdvisor, a tool
designed to assist users in defining their HPC clusters in the cloud. It
considers the application's input and utilizes a major cloud provider as
a use case for its back-end component.

\end{abstract}


\section{Introduction}

HPC cloud has become a reality in several industries. When companies and
research institutions began assessing the feasibility of using the cloud for HPC
workloads, extensive research was conducted on performance evaluation and cost
analysis~\cite{netto2018hpc}. Today, we see various ways HPC is being utilized
in the cloud, especially with the rise of new AI-based applications. Companies
from multiple industries are creating Software-as-a-Service solutions to host
simulators. Some companies provision on-demand multi-user clusters, offering
them as Infrastructure-as-a-Service. Additionally, institutions provide users
with cloud credits, enabling them to create their own clusters or resource
pools on-demand.

A typical HPC application requires instructions on where to run, including the
resource type, number of cluster nodes, and processes per node. In an
on-premises setting, the resource type may correspond to a queue defined by
a scheduler. In a cloud environment, the variety of resource types is often
greater compared to on-premises resources due to the flexibility of configuring
resources and the faster availability of new options. Traditionally,
benchmarking is conducted to determine the best cost-performance setup.
However, not every group or institution may have the resources or expertise to
perform this benchmarking.

In this context, a user intending to run an application with specific input
parameters or files would ideally use a service or tool to specify these
details. The tool would then provide a list of resource options, considering
performance and cost trade-offs, to assist in making the final decision. With
a substantial database of historical executions and an application with
a reduced set of input parameters that influence resource selection, it may be
possible to generate this list of resource options without the need for
additional testing or execution. In a recent paper~\cite{lamar2023evaluating},
Lamar \textit{et al.}  discuss job run time predictions considering application input
parameters, and they highlight that although applications have a wide variety
of parameters, a few of them actually influence application resource
consumption/execution times.

Moving to the cloud or experimenting with new workloads means that such data is
not always available or complete. Generating such data means creating an environment in the cloud, generating
scenarios, executing the scenarios, collecting the data, deactivating the
environment, and filtering and organizing the data. These steps may be time-consuming, especially when one is in the early stages of moving HPC workloads to
the cloud.

In this paper we introduce HPCAdvisor, an open-source tool to advise users on
cloud resource selection decision for HPC workloads. The tool aims at helping
users select VM type, number of nodes, and processes per node taking into
account application input. Its current implementation focuses on the automated
data collection given an easy-to-use, yet flexible, interface to users. With
basic information such as Cloud account, specifications of scenarios defined by
VM types, number of nodes, processes per node, application input parameters,
and the specifications to set up and run the application, the entire cloud
environment is automatically created, scenarios are executed with all
combinations of user input, data is collected, filtered, and organized. Users
can also access automatically generated plots that expose data on execution time,
costs, speed up, and efficiency. Advice is provided as a Pareto front form,
exposing the best solutions in the search space considering execution
time and cost.

The contribution of the paper is a description of the end-to-end design,
implementation, and user experience of the tool to achieve a fully automated
life cycle for data collection to assist in resource selection for HPC
workloads in the cloud. We present a few examples of automatically generated
outputs using both OpenFOAM and LAMMPS applications with up to two thousand
cores and VMs with InfiniBand networks. We also discuss ongoing optimizations we
are implementing to minimize the need to run certain scenarios based on
existing runs. Its implementation is available on
GitHub\footnote{\url{https://azure.github.io/hpcadvisor/}} and contains
examples for well-known HPC applications, such as WRF, NAMD, OpenFOAM, LAMMPS,
and GROMACS. We describe the implementation relying on a back-end based on Azure Batch, which is a middleware to support cloud-native executions of various workloads in Azure. This back-end can be replaced if needed.

\bigskip

In its current format the benefits of the tool are:

\begin{itemize}
  \item \textbf{Automated data collection:} The tool automatically creates the cloud environment, executes scenarios, collects data, and organizes it.
  \item \textbf{Automated plots:} The tool generates plots that expose data on execution time, costs, speed up, and efficiency.
  \item \textbf{Flexible:} The tool relies on bash scripts to set up and run applications, which can be easily customized.
  \item \textbf{Parameter Sweep:} As is, the tool can be used to perform parameter sweeps for HPC applications in a cloud environment, providing a full end-to-end solution.
  \item \textbf{Advice:} The tool provides advice in Pareto front form, exposing the best solutions in the search space considering execution time and cost.
  \item \textbf{Open source:} The tool is open source and can be expanded according to user needs.
\end{itemize}

\bigskip

In its future format, we plan to:
\begin{itemize}
\item \textbf{Smart sampling:} Implement a module that can be used as a stand-alone to minimize the need to run certain scenarios based on existing runs. Having this module as a stand-alone allows its usage in situations where there are already existing tools in place for resource provisioning, job submission, and/or benchmarking.
\item \textbf{Comprehensive advice:} Implement a module to provide more comprehensive advice; e.g., apart from providing only the Pareto front, we envision the advice being used to provide recipes to run jobs (e.g., Slurm scripts) or computing environment creation/modification (e.g., cluster creation or scheduling queue creation/modification).
\end{itemize}

\begin{figure*}[!ht]
        \centering
       \includegraphics[width=1.0\textwidth]{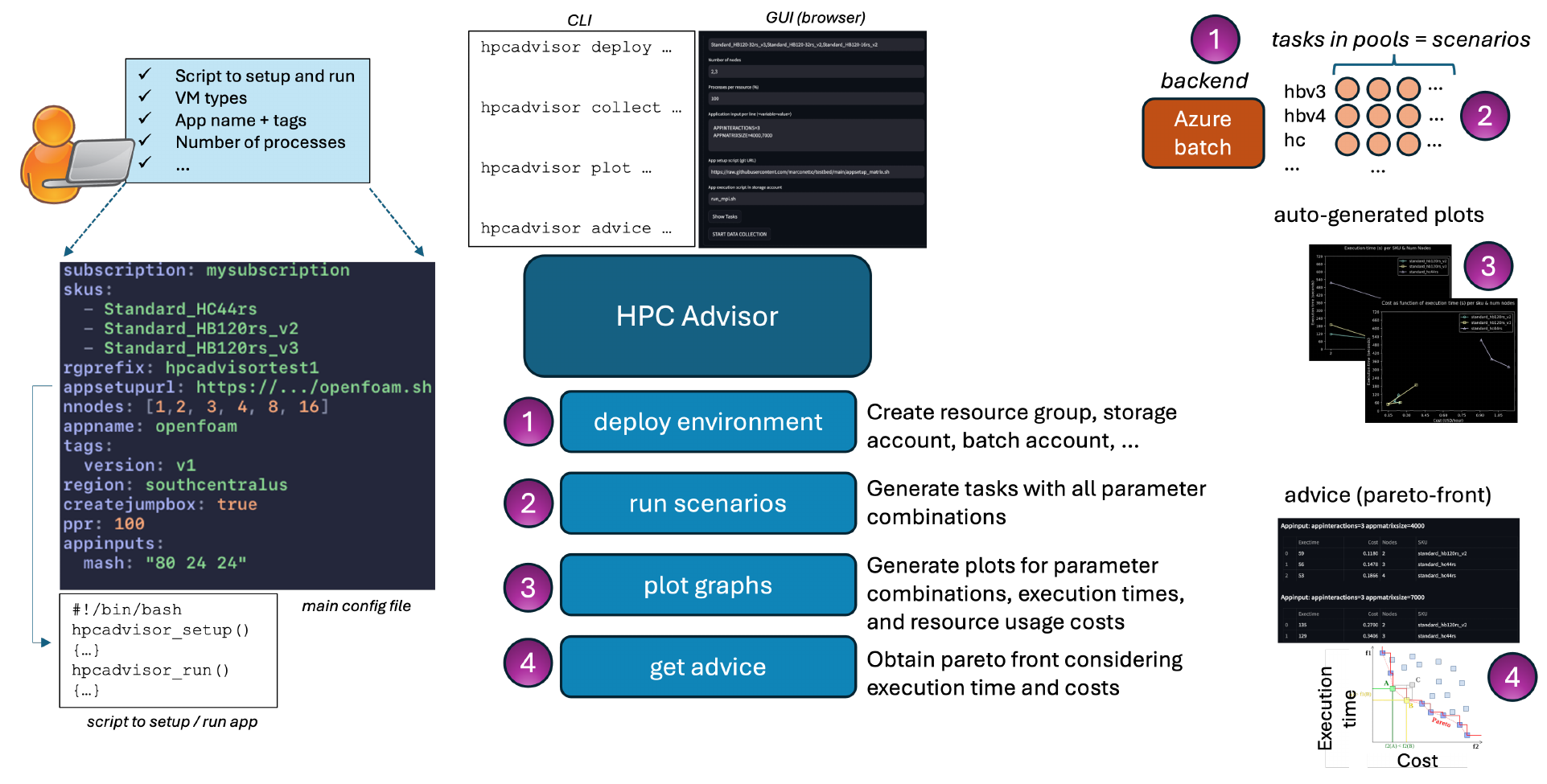}
        \vspace{-10mm}
        \caption{Overview of the HPCAdvisor tool.}
        \label{fig:architecture}
\end{figure*}

\section{Related Work}

Related to our work are efforts on (i) application performance prediction
models and tools; (ii) cloud resource selection; and (iii) HPC benchmarking tools.

Application performance prediction has been vastly studied over the
years~\cite{flores2021performance,wang2019novel,tsafrir2007backfilling,yang2005cross,smith2007prediction,lamar2023evaluating,mariani2018predicting,betting2023oikonomos},
both for HPC and non-HPC applications. Performance prediction helps with capacity
planning, including cases where different hardware configurations need to be
assessed for new resource acquisition. Several companies as well as research
and engineering institutions have benchmarking teams to help with this task.
Another driving force for performance prediction studies is job scheduling.
Traditionally, job schedulers require users to estimate runtime of their jobs
in order to make scheduling decisions. Such estimates are hard for end-users to obtain~\cite{lee2006user,bailey2005user}.

Several techniques have been explored for performance prediction. For instance,
motivated by improving scheduling outcomes, Tsafrir \textit{et al.}~\cite{tsafrir2007backfilling} developed a technique for job scheduling
based on system-generated job runtime estimates that uses the average time of
the previous two job runtime values. This strategy comes from their conclusion
that users tend to submit similar jobs over a short period of time. Yang \textit{et al.}~\cite{yang2005cross}'s technique predicts the execution time of jobs across
multiple platforms. They use data collected from short executions of a job and
the relative performance of each platform. After warm-up phases, several
computer simulations may have fixed execution time per time step.
Smith~\cite{smith2007prediction} developed a prediction system based on
Instance-Based Learning techniques, leveraging genetic algorithms to refine
input parameters of the prediction model. Yang \textit{et al.}~\cite{yang2023exploring}
introduce a framework called PREP (Path RuntimE Prediction), which explores the
running path of jobs to predict their runtime. Their framework uses support
vector regression, decision trees, and random forest. Lamar
\textit{et al.}~\cite{lamar2023evaluating} conduct a study on job run time prediction
focusing on application input parameters. They evaluated over 20 machine
learning model variants and analyzed the influence of the various input
parameters for multiple applications, including ExaMiniMD, LAMMPS, NEKbone,
SWFFT, and HACC.

We have also seen efforts on performance prediction in the context of cloud
computing. For instance, Mariani \textit{et al.}~\cite{mariani2018predicting} introduce
a machine-learning methodology to help users select the best cloud configurations
for a given workload before deploying it in the cloud. They couple
a cloud-performance-prediction model on the cloud-provider side with
a hardware-independent profile-prediction model on the user side. Users need to
do some profiling on their applications using their on-premises resources.
Betting \textit{et al.}~\cite{betting2023oikonomos} introduce Oikonomos, a data-driven
resource-recommendation system for HPC workloads in the cloud. Their system is
based on a Multi-layer Perceptron to predict application performance by exploring
both application input parameters and instance types. Brunetta and
Borin~\cite{brunetta2019selecting} propose a methodology to help users select
cloud resources for HPC workloads. Their predictions rely on the first
paramount iterations of the application for each cloud resource selection.
They highlight the importance of using actual input of the application versus
smaller input datasets. Samuel \textit{et al.}~\cite{samuel2020a2cloud} introduce
a framework called A2Cloud-RF, which aims at matching scientific applications to
cloud instances. They rely on a random-forest model, which allows predictions based
on input application's characteristics or application classes.

On the HPC benchmarking side, there are several tools and services available.
Here we highlight a few of them. For instance, ReFrame~\cite{karakasis2020enabling} is a framework for system regression tests and benchmarks focused on HPC environments. It allows users to write portable tests in a declarative way and hides complexity of several stages of the tests including  compilation, job submission, and, programming environment, among others. Ramble~\cite{ramble2024} is a multi-platform framework that can be used to configure experiments, including scientific parameter sweeps,
performance scaling studies, and compiler flag sweeps. Pavilion2~\cite{pavilion2} is a framework for running and analyzing tests on HPC systems. Users can specify experiments using YAML files and the framework hides the complexity of interacting with the underlying systems to generate the benchmarking results.

We aim to deliver an easy-to-use, practical, flexible, and open-source tool for
HPC cloud resource selection, incorporating existing and cutting-edge
prediction techniques throughout the development process.

\section{HPCAdvisor: Design and Implementation}

In this section we explain in detail the design and implementation of the
HPCAdvisor tool. We provide an overview of the tool, the major configuration files and
scripts, the internal flow of execution, and how the tool interacts with
cloud services. We also describe how the tool can be expanded and some
ongoing developments. The tool is open source and in constant development.

Figure~\ref{fig:architecture} depicts an overview of the HPCAdvisor tool. The
core HPCAdvisor tool is a Python-based program that is invoked by the end-user.
The tool can be used in the browser or Command Line Interface (CLI). Based
on user input, HPCAdvisor automatically provisions the cloud environment to run
different scenarios, collects all data from such executions, and deletes the
cloud environment if user desires. Once this is done, users can generate plots
and get advice data to make the resource selection based on a Pareto front of
the best options considering execution times and costs.  Users may choose to run all
scenarios, or rely on module to do smart sampling, i.e. it selects a subset of
scenarios based on existing data from previous executions. We will go into
further details of all these steps in the next sections.

\subsection{User Input}
\label{sec:userinput}

The tool expects two major input files from the user. The first one is the main
YAML configuration file, which requires the following input:

\begin{itemize}
    \item \textbf{Cloud subscription:} ID or name of the cloud subscription where all resources are provisioned;
    \item \textbf{Resource group prefix:} string used as a prefix of the resource groups where all resources are provisioned;
    \item \textbf{Region}: geographical region where the cloud resources are provisioned;
    \item \textbf{Application setup url:} URL that specifies how the application is set up and executed;
    \item \textbf{Processes per resource:} percentage of processes per resource;
    \item \textbf{Application inputs:} application inputs (e.g., matrix size for the matrix multiplication application, number of cells or mesh definition for a CFD application such as OpenFOAM, resolution for a weather forecast such as WRF);
    \item \textbf{VM types / SKUs:} specifies the list of VM types that user wants to test;
    \item \textbf{Number of nodes:} list of the number of nodes that user wants to test, e.g. 1, 2, 4, 8, 16, 32;
    \item \textbf{Tag:} identifications to be included into the results of the experiments.

\end{itemize}

There are also a few optional parameters that are related to VPN and jumpbox. It is a common practice to provision resources with private IP addresses. For instance, a user can set up a VPN and peer it to the HPCAdvisor virtual network. HPCAdvisor also allows the creation of a jumpbox machine. This machine can, for instance, have access to the shared file system used by the cluster nodes, allowing the user to log in and easily see all files being created by the multiple scenarios under execution:

\begin{itemize}
    \item \textbf{VPN resource group (optional):} existing resource group that contains a VPN setup;
    \item \textbf{VPN virtual network (optional):} existing vnet name for the VPN setup;
    \item \textbf{Peer VPN (optional):} boolean for peering with VPN resource group / vnet;
    \item \textbf{Create jumpbox (optional):} boolean for creating a VM in the same resource group.
\end{itemize}

Listing~\ref{lst:mainconf} illustrates an example of this main configuration file. This example represents an experiment for OpenFOAM, which assesses three VM types, six number of nodes, and two definitions of mesh sizes. This generates 3x6x2 scenarios.

\begin{figure}[hb]
  \centering
\begin{lstlisting}[language=yaml, caption=HPCAdvisor main user configuration file., label={lst:mainconf}, linewidth=\columnwidth]
# Example of main configuration file

subscription: mysubscription
skus:
  - Standard_HC44rs
  - Standard_HB120rs_v2
  - Standard_HB120rs_v3
rgprefix: hpcadvisortest1
appsetupurl: https://.../openfoam.sh
nnodes: [1, 2, 3, 4, 8, 16]
appname: openfoam
tags:
  version: v1
region: southcentralus
createjumpbox: true
ppr: 100
appinputs:
  mesh: "80 24 24"
  mesh: "60 16 16"
\end{lstlisting}
\end{figure}

The other main input from the user is the specification of (i) how to set up the
application and (ii) how to run the job, which is a bash script containing these two
major functions. Here, job is the application plus its input parameters to be
executed. The setup of the application has two major parts: (a) download of
input data and (b) preparation of the application (e.g. download application,
compile application, setup licenses, or any other mechanism to have the
application ready for execution).

\medskip
\noindent \textbf{Application setup.} There are various ways in which an
application becomes available to the end user. It can be via downloading
a binary file, through a set of compilation scripts, or through an
application installation framework such as spack~\cite{gamblin2015spack},
easybuild~\cite{hoste2012easybuild}, and EESSI~\cite{droge2023eessi}. The
application setup script can also be used to download data for the application.
This script is called every time a new VM type starts to be tested. If the user
does not want to prepare the application again, or download the data again,
a simple test can be done to avoid repeating such setup.

\medskip
\noindent \textbf{Application run.} This is where one specifies how the
application runs. This can be a simple \texttt{mpirun} command, or it may contain
other tasks including,  preparation of input files based on
environment variables,  setting up of process pinning, parse of application metric
data to be exposed to the generation of plots or advice, among others.

\lstset{escapeinside={(*@}{@*)}}
\begin{figure}[!h] 
  \centering
\begin{lstlisting}[language=bash, caption=Application setup/run example for LAMMPS, label={lst:appsetup}, linewidth=\columnwidth]
#!/usr/bin/env bash

(*@\textbf{hpcadvisor\_setup()}@*) {

  if [[ -f in.lj.txt ]]; then
    echo "Data already exists"
    return 0
  fi

  wget https://www.lammps.org/inputs/in.lj.txt
}

(*@\textbf{hpcadvisor\_run()}@*) {

  source /cvmfs/software.eessi.io/versions/2023.06/init/bash
  module load LAMMPS

  inputfile="in.lj.txt"
  cp ../$inputfile .

  sed -i "s/variable\s\+x\s\+index\s\+[0-9]\+/variable x index $BOXFACTOR/" $inputfile
  sed -i "s/variable\s\+y\s\+index\s\+[0-9]\+/variable y index $BOXFACTOR/" $inputfile
  sed -i "s/variable\s\+z\s\+index\s\+[0-9]\+/variable z index $BOXFACTOR/" $inputfile

  NP=$(($NNODES * $PPN))
  export UCX_NET_DEVICES=mlx5_ib0:1
  APP=$(which lmp)
  mpirun -np $NP --host "$HOSTLIST_PPN" "$APP" -i $inputfile

  log_file="log.lammps"

  if grep -q "Total wall time:" "$log_file"; then
    echo "Simulation completed successfully."
    APPEXECTIME=$(cat log.lammps | grep Loop | awk '{print $4}')
    LAMMPSATOMS=$(cat log.lammps | grep Loop | awk '{print $12}')
    LAMMPSSTEPS=$(cat log.lammps | grep Loop | awk '{print $9}')
    echo "HPCADVISORVAR APPEXECTIME=$LAMMPSCLOCKTIME"
    echo "HPCADVISORVAR LAMMPSATOMS=$LAMMPSATOMS"
    echo "HPCADVISORVAR LAMMPSSTEPS=$LAMMPSSTEPS"
    return 0
  else
    echo "Simulation did not complete successfully."
    return 1
  fi
}


\end{lstlisting}
\end{figure}

Listing~\ref{lst:appsetup} shows a full script to set up and run LAMMPS in the HPCAdvisor tool. As we can see here we have two functions, \texttt{hpcadvisor\_setup} and \texttt{hpcadvisor\_run} that sets up and runs the job, respectively. This example uses EESSI to access the LAMMPS application. Therefore, the setup of the application consists of only downloading the input data. The second function contains a few more instructions. It starts by enabling the use of EESSI and loading the LAMMPS module (Lines 17--18), which as a consequence loads \texttt{mpirun} as well---so both the application and the MPI environment are provided by EESSI. Then it copies the download file from the parent directory to the local directory that the script runs (Line 21). Every job contains its own directory which is automatically created by HPCAdvisor. Next step is to update the input file according to the application input parameter we want to assess. In this case, we want to use a box multiplication factor to increase the number of atoms in the simulation (Lines 23--25). Once the input is ready, we specify a few variables for the \texttt{mpirun} command to be executed (Lines 27--30). This part could be extended to exercise different CPU pinning strategies for example. The last part (Lines 32--46) has two major goals. One is to assess if the application concluded its execution properly, and one way to do is to check the content of its log file. The second goal is to obtain some metrics from the output, which will then be exposed into the dataset file, which contains all output of the scenarios that were executed. Thus, any line containing \texttt{"HPCADVISOR variable=value"} is saved in the dataset file.

There are a few relevant variables that can be used inside the \texttt{hpcadvisor\_run} function. Table~\ref{tab:variables} describes those variables.

\begin{table}[!h]
\centering
\caption{Environment variables.}
\label{tab:variables}
\begin{tabular}{ l p{5cm} }
\hline
\hline
\textbf{Variable} & \textbf{Description} \\
\hline
\texttt{NNODES} & Number of cluster nodes \\
\texttt{PPN} & Processes per node \\
    \texttt{SKU}, \texttt{VMTYPE} & Virtual machine type \\
    \texttt{HOSTLIST\_PPN} & List of hosts and their PPN \\
\texttt{HOSTFILE\_PATH} & Path of hostfile \\
\texttt{TASKRUN\_DIR} & Directory of the job run \\
\hline
\end{tabular}
\end{table}

\subsection{Environment Deployment}

To collect new data for application executions with different input parameters,
a cloud environment needs to be deployed. HPCAdvisor is able to automatically
create a cloud environment to collect such data. In the current
implementation, HPCAdvisor's back-end relies on Azure, as the cloud provider, and
Azure Batch, as the resource orchestrator.

Here is the sequence used to provision the resources:

\begin{enumerate}
\item \textbf{Variables.} First step is to set up variables to be used in all the other steps, including names for the resources, images to be used, etc.
\item \textbf{Basic landing zone.} Next, a resource group is created, along with a virtual network and subnet.
\item \textbf{Storage account.} Then a storage account to store batch-related files and NFS is created.
\item \textbf{Batch service.} The final step is to create a batch service with no resources.
\item \textbf{Jumpbox and network peering.} Optionally, HPCAdvisor can also create a jumpbox machine and perform network peering, for instance, when a VPN is in another virtual network. The jumpbox can be useful for logging in and easily navigating files created during the scenario executions or performing small tests.
\end{enumerate}

As HPCAdvisor is open source, the back-end can be replaced. We plan to create
a couple of other back-end examples, including one that uses Slurm directly.

\subsection{Data collection}

Once the environment is deployed, the data collection phase can begin. The
first step is to create the list of scenarios (or tasks) to be executed based
on the main configuration file (Listing~\ref{lst:mainconf}). Here we take all
the VM types, number of nodes, processes per node, and application input
parameters to generate all combinations. This list is recorded and stored in
a JSON file. The list also contains the status of the task, which can be
pending, failed, or completed.

Then, a loop goes through the list of tasks to execute them and
collect data as described in
Algorithm~\ref{alg:process_tasks}. Every time a new VM type
needs to be assessed, a new pool of resources is created, and
a setup task is executed. The number of nodes that the user
requested for testing is then incremented in the pool. Once the
pool is utilized, it is reduced to zero nodes or deleted, depending
on user preference.

\bigskip

\noindent \textbf{Costs for data collection.} It is important to note that data collection incurs a cost. In a later section, we will describe approaches we are investigating to reduce the number of scenarios that need to be executed. From a cost perspective, users typically do not collect data solely to obtain advice for a single production execution. Instead, they often perform parameter sweeps, leading to multiple executions with similar resource usage patterns, which helps offset the cost of the advice. When this payoff occurs depends on the application, its input parameters, the number of scenarios executed, and the resource usage.

\begin{algorithm}[h]
    \caption{Process a List of Tasks (Scenarios)}
\label{alg:process_tasks}
\KwIn{A list of tasks, \textit{tasks}}
\KwOut{Processed results}

\textit{previousVMType} $\gets \emptyset$\;

\ForEach{\textit{task} \textbf{in} \textit{tasks}}{
    \If{\textit{previousVMType} $\neq$ \textit{task.vmtype}}{
        \If{\textit{pool} exists}{
            resize pool to zero or delete pool\;
        }
        create\_setup\_task(\textit{task})\;
    }
    pool $\gets$  resize\_pool(\textit{task.vmtype}, \textit{task.nnodes})\;
    create\_compute\_task(\textit{task})\;
    execute\_compute\_task(\textit{task})\;
    store\_task\_data(\textit{task})\;
    update\_task\_status(\textit{task}, completed)\;
    \textit{previousVMType} $\gets$ \textit{task.vmtype}\;
}
\If{\textit{pool}}{
     resize pool to zero or delete pool\;
}
\Return{Processed results}

\end{algorithm}

\subsection{Plots}

Once the data is recorded from the data collection phase, plots and advice can be created.

For now, HPCAdvisor generates four types of plots:

\begin{enumerate}
    \item \textbf{Execution Time vs Number of Nodes.} Plots the execution time as a function of the number of nodes for each VM type.
    \item \textbf{Execution Time vs Cost.} Plots the execution time as a function of the cost to run each task for each VM type.
    \item \textbf{Speed up.} Plots how much faster the parallel execution over multiple nodes is compared to single node.
    \item \textbf{Efficiency.} Plots how effectively the computing resources are being utilized when solving a problem using multiple nodes (speed up over number of nodes).
\end{enumerate}

\begin{figure}[!ht]
    \centering
    \includegraphics[width=0.80\columnwidth]{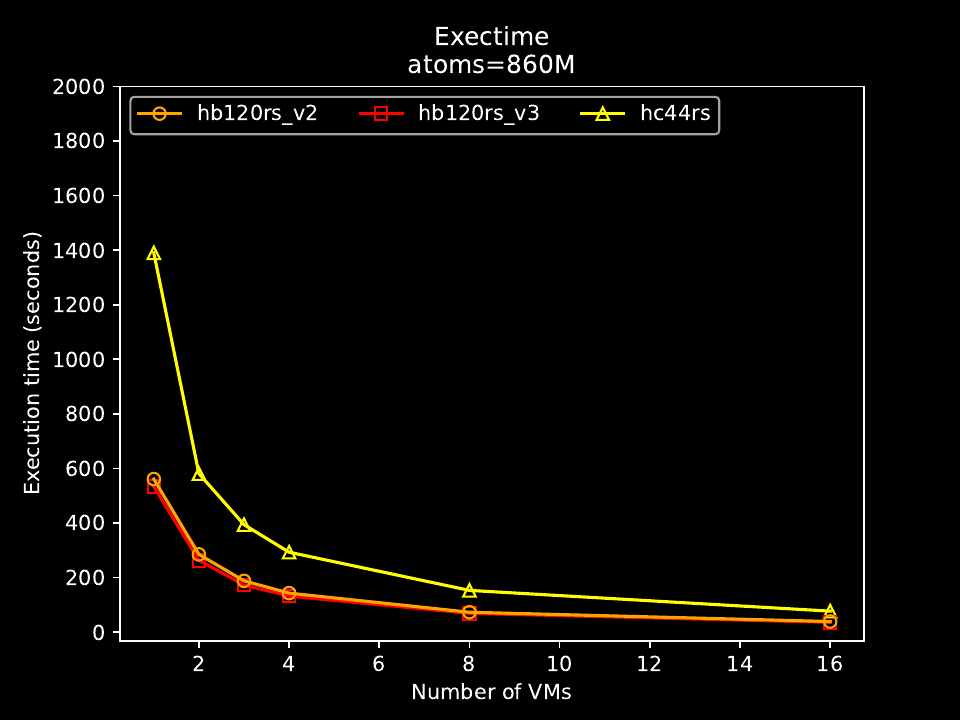}
    \caption{Plot example: Execution Time vs. Number of Nodes.}
    \label{fig:plotexample0}
\end{figure}

\begin{figure}[!ht]
    \centering
    \includegraphics[width=0.80\columnwidth]{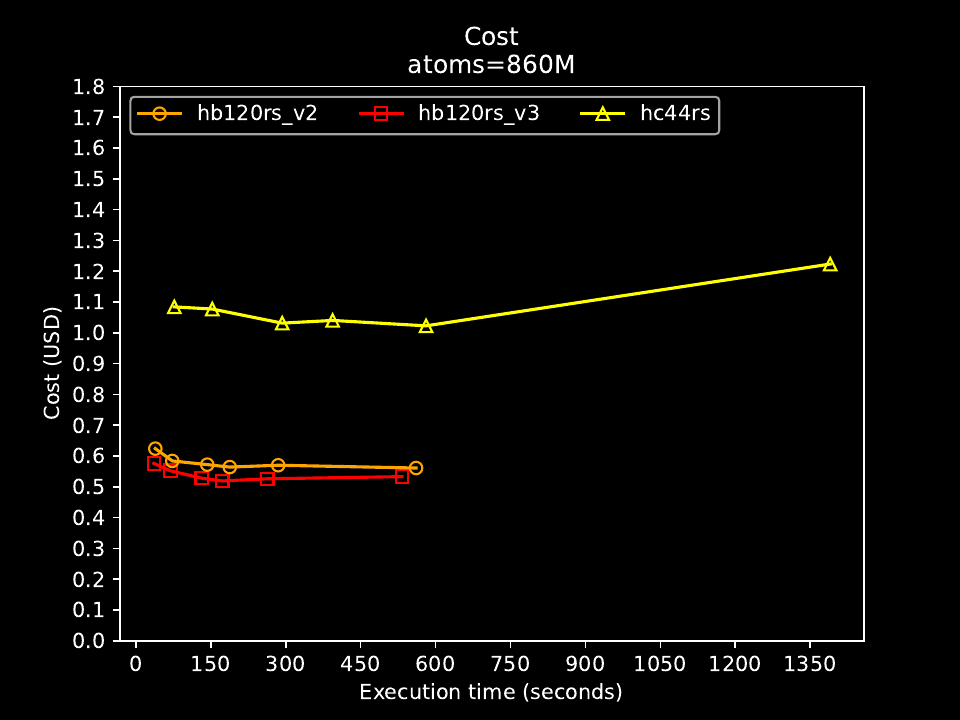}
    \caption{Plot example: Execution Time vs. Cost.}
    \label{fig:plotexample1}
\end{figure}

\begin{figure}[!ht]
    \centering
    \includegraphics[width=0.8\columnwidth]{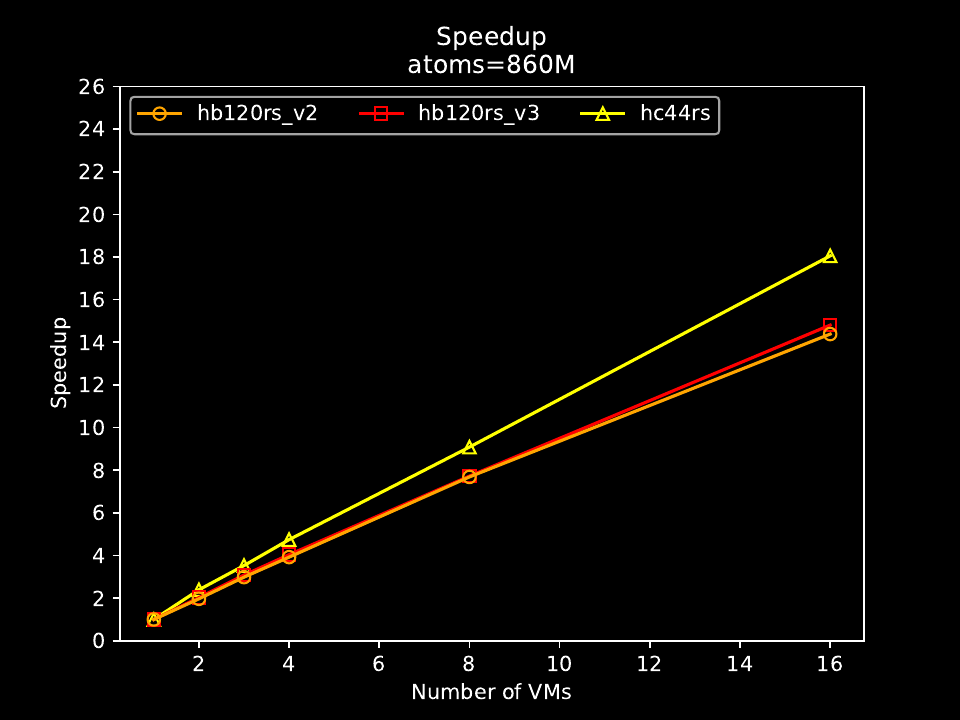}
    \caption{Plot example: Speed up.}
    \label{fig:plotexample2}
\end{figure}

\begin{figure}[!ht]
    \centering
    \includegraphics[width=0.8\columnwidth]{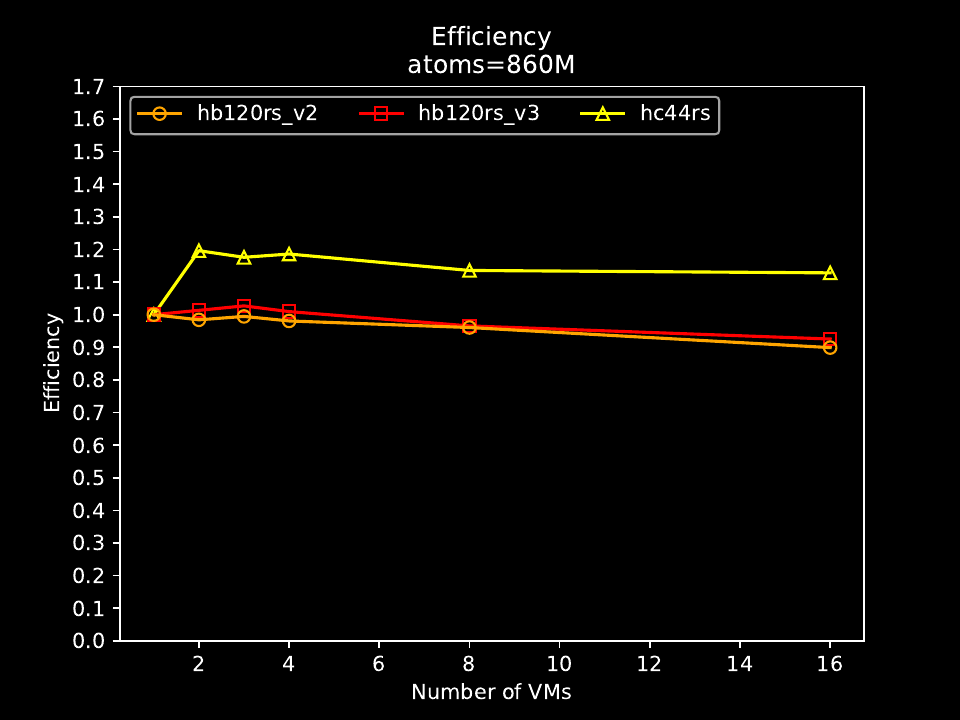}
    \caption{Plot example: Efficiency.}
    \label{fig:plotexample3}
\end{figure}

For the plot feature, we also allow customization of subtitles for the plots.
Here are the plots exactly as created by the HPCAdvisor tool (Figures
\ref{fig:plotexample0}, \ref{fig:plotexample1}, \ref{fig:plotexample2},
\ref{fig:plotexample3}). This is an example for a LAMMPS workload with three VM
types, each containing 44, 120, and 120 cores, interconnected by an InfiniBand network. Scenarios run up to 1,920 cores. When using
the CLI, the plots are generated in the current folder, and when using the GUI, the
plots are presented in the UI to the user. The cost represented here is for the VMs only, without considering other costs such as software license, storage, or any additional services. In Figure \ref{fig:plotexample3}, we observe an efficiency greater than 1, which represents a super linear speed up using multiple nodes.

\subsection{Advice}

\begin{figure}[!h]
\centering
\begin{tikzpicture}
    \begin{axis}[
        xlabel={Cost},
        ylabel={Execution Time},
        grid=both,
        legend pos=north east,
        width=9cm,
        height=6cm,
    ]
    \addplot[
        color=blue,
        mark=o,
        only marks,
        mark size=3pt,
    ]
    coordinates {
        (0.2, 1.0)
        (0.8, 0.8)
        (0.4, 0.4)
        (0.4, 0.2)
        (0.4, 0.3)
        (0.3, 0.3)
        (0.3, 0.7)
        (0.2, 0.5)
        (0.53, 0.87)
        (0.65, 0.67)
        (0.8, 0.6)
        (0.8, 0.6)
        (0.85, 0.65)
        (0.8, 0.2)
        (0.8, 0.15)
    };
    \addlegendentry{Scenarios}

    \addplot[
        color=red,
        mark=triangle*,
        thick,
        smooth,
        mark options={solid, fill=none},
    ]
    coordinates {
        (0.2, 0.5)
        (0.3, 0.3)
        (0.4, 0.2)
        (0.8, 0.15)
    };
    \addlegendentry{Pareto Front}

    \end{axis}
\end{tikzpicture}
    \vspace{-4mm}
    \caption{Advice based on pareto front}
    \label{fig:pareto}
\end{figure}
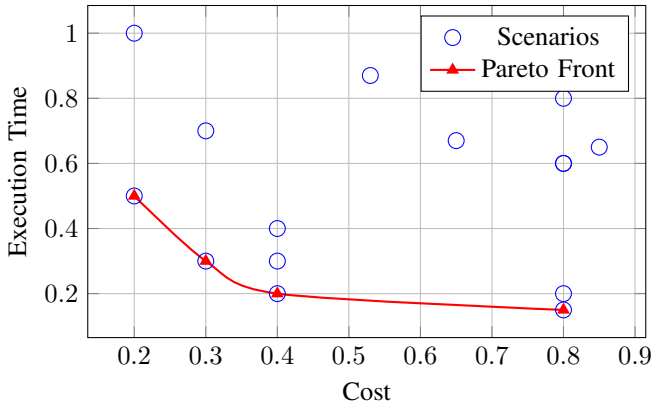

When providing advice on how a given workload should be run, the two major
factors usually considered are performance (time to completion) and costs (money
spent to run the workload).
Sometimes, a more efficient execution may cost more
money, while a cheaper execution may take longer. Additionally, there are
cases where the best option is indeed the cheapest and fastest. For this
reason, providing the user with a single option to run a given workload may not
be the best approach. Even if we let users specify their main objective,
i.e. cost or performance, the choice may still be subjective. For instance,
would one be willing to get 20\% faster results by paying 10\% more, or
a solution that is 18\% faster by paying 2\% more? For this reason, providing
the Pareto front that considers both execution time and costs seems to be
a better approach.  The Pareto front represents the solutions that
are Pareto efficient, i.e. a set of solutions that are non-dominated relative to each
other but are superior to the rest of solutions in the search space.
Figure~\ref{fig:pareto} helps illustrate the concept; execution time
and cost are the factors we want to optimize (Y axis and X axis); blue circles represent
the executed scenarios and the red line represents the Pareto front.

The tables below (Listing~\ref{lst:advice} and Listing~\ref{lst:advice2})
represent the actual advice output from HPCAdvisor for two examples: (i) the OpenFOAM
example with the configuration of ``BLOCKMESH\_DIMENSIONS" set to ``40 16 16"
for the motorBike example containing 8 million cells and (ii) the LAMMPS
example, which is based on the official LAMMPS Lennard Jones benchmark (“atomic
fluid with Lennard-Jones potential”), in which we multiply the box dimensions
by 30 to obtain 800 million atoms. The advice data presented here is sorted by the least execution time first, but the tool has the option to have the data sorted by cost as well.

\begin{figure}[!h]
  \centering
    \begin{lstlisting}[language=yaml, caption=Example advice: OpenFOAM., label={lst:advice}, linewidth=\columnwidth]
Exectime(s)       Cost($)       Nodes       SKU
34                0.5440        16          hb120rs_v3
38                0.3040        8           hb120rs_v2
48                0.1920        4           hb120rs_v3
59                0.1770        3           hb120rs_v3
\end{lstlisting}
\end{figure}

\begin{figure}[!h]
  \centering
    \begin{lstlisting}[language=yaml, caption=Example advice: LAMMPS., label={lst:advice2}, linewidth=\columnwidth]
Exectime(s)        Cost($)       Nodes       SKU
36                 0.5760        16          hb120rs_v3
69                 0.5520        8           hb120rs_v3
132                0.5280        4           hb120rs_v3
173                0.5190        3           hb120rs_v3
\end{lstlisting}
\end{figure}

\subsection{Optimizations for scenario generation and executions}

Another key component of the HPCAdvisor, which is still under development, is
the optimization of scenario generation and execution. We envision a solution
to resource selection in which a user would provide the application with its
input files and parameters, and the user would receive a list of options (e.g. the
Pareto front discussed previously) to run their workloads, and this list
would require minimal or no executions in the cloud.

\begin{table*}[!h]
\centering
\caption{Commands for CLI execution mode.}
\label{tab:cli}
\begin{tabular}{|c|c|p{10cm}|}
    \hline
    \textbf{Command} & \textbf{Subcommand} & \textbf{Description} \\
    \hline
     & \texttt{create} & Creates a cloud deployment \\
    \texttt{deploy} & \texttt{list} & Lists all previous and current cloud deployments. \\
     & \texttt{shutdown} & Shuts down a given cloud deployment, deleting all its resources. \\
    \hline
    \texttt{collect} & \texttt{-} & Collects data, i.e. runs all scenarios on a given deployment. \\
    \hline
    \texttt{plot} & \texttt{-} & Generates plots using a given data filter.\\
    \hline
    \texttt{advice} & \texttt{-} & Generates advice (i.e. Pareto front) using a given data filter. \\
    \hline
    \texttt{gui} & \texttt{-} & Starts the GUI mode. \\
    \hline
\end{tabular}
\end{table*}

If there is enough data from previous executions, depending on the application,
it may be possible to create a machine learning-based model (existing
literature shows some efforts in this
area~\cite{lamar2023evaluating,mariani2018predicting,samuel2020a2cloud}). In
certain scenarios with small amounts of data, a simple regression analysis
could help, especially if the application is better-behaved. If
scenario execution is required, we can then also explore the fact that several HPC
workloads have a steady execution time per step (after warm-up)~\cite{yang2005cross,brunetta2019selecting}. So one could get some
approximation of execution times and costs.

Our focus now on optimizations is to assume that little or no data is
available. Therefore, the strategy we are investigating is to have a flexible
set of scenarios to be executed at the data collection phase. The strategy
would then be to identify which new scenarios would need to be executed to
obtain the best ``return on investment'', i.e. scenarios that would help
provide more information for generating the Pareto front. It is worth
highlighting that our aim is not to determine the exact execution times and
costs for all scenarios, but to generate a Pareto front to advise the user
on resource selection. We want to avoid using computing resources to find
information in a search space; problem that can be mapped to Design of
Experiments.

A few strategies in this direction we are currently investigating are:

\begin{itemize}

\item \textbf{Aggressive scenario discarding.} Whenever there is evidence, at
    a given threshold, that a VM type will probably not be part of the Pareto
        front, we ignore all scenarios with that VM type.

\item \textbf{Fixed performance factor.} Some applications scale well, so by
    identifying the influence of the application input parameters and using the
        data from previous scenarios, new curves could be
        identified. We are currently exploring regression techniques and obtaining positive results for some workloads. For instance, by using the same VM type but different application input parameters and their influence on execution time, or by using the same application input parameters but analyzing a different VM type, we can identify scenarios that should or should not be in the Pareto front.

\item \textbf{Infrastructure bottlenecks}. For each scenario executed, it is
    possible to identify how long it took. However, with proper monitoring, it
        is also possible to identify possible bottlenecks while executing the
        scenario via infrastructure related metrics such as CPU, memory,
        network utilization. This can also serve as a hint to identify and prioritize the
        next scenarios to be executed, or even discarding ones that will not be part of the Pareto front.

\end{itemize}

\section{Tool Usage}

The setup and usage of the HPCAdvisor tool is straightforward. The user needs
to clone the Git repository and activate the Python virtual environment or
install the tool via \texttt{PIP}. Once this is done, the tool can be used
either via the Graphical User Interface (GUI) (browser) or via the command line interface (CLI).

\begin{figure}[!ht]
    \centering
    \includegraphics[width=1.0\columnwidth]{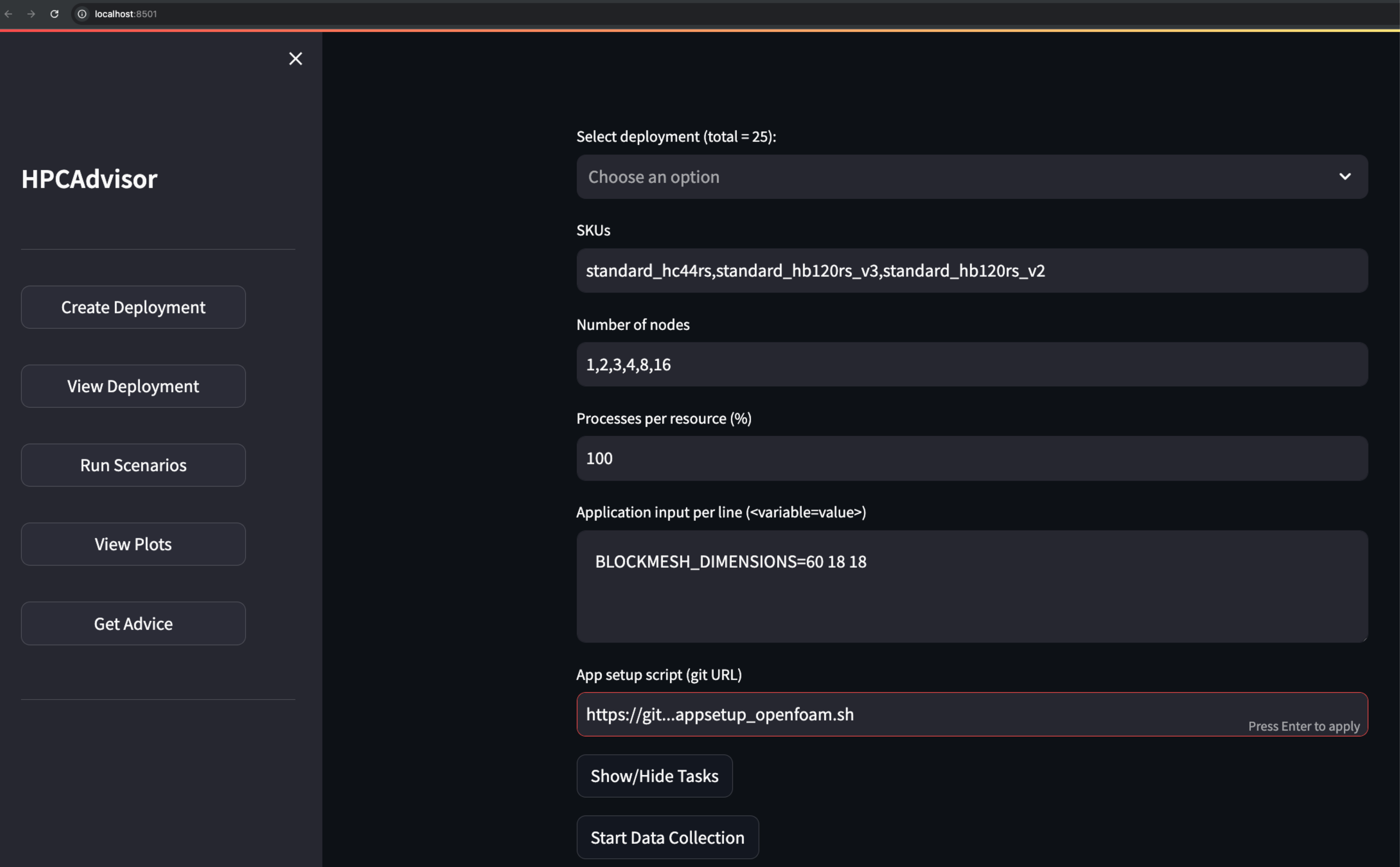}
    \caption{Screenshot of the tool in the data collection step.}
    \label{fig:gui}
\end{figure}

Figure~\ref{fig:gui} shows a screenshot of the tool under the GUI execution
mode. On the left side of the screen, we have the major operations that the tool
can perform, and on the right side, we see the screen of the data collection
phase, where a deployment is already available in the cloud, and the user can
start to collect data.

The user can also opt for a Command Line Interface (CLI) execution, where commands and subcommands are available, as described in Table~\ref{tab:cli}.
The major input files are the two ones described in Section~\ref{sec:userinput}.

\section{Final Remarks and Next Steps}

This paper introduced the vision and current development status of a tool to
help users select cloud resources to run HPC workloads. The tool is open
source, which allows it to be further extended according to different environments and
needs.

We believe the current implementation of the tool is flexible enough to support custom definitions for setting up an application to run, while keeping the user input simple and intuitive. The tool, in its current form, serves as an easy mechanism for running benchmarks in the cloud. We have made significant progress in generating data, visualizing results through plots, and providing users with advice on resource selection, taking into account both execution time and cost. We have successfully tested it with applications such as WRF, OpenFOAM, GROMACS, LAMMPS, and NAMD.

The ongoing effort for the tool is to focus on optimizing the necessary
executions to obtain such advice while expanding the examples to consider
applications with multiple input parameters and various ways they impact the execution time and the consumption of computing resources. We also want to ensure the user
experience using the tool continues to be straightforward with no additional
setups, which is particularly important for users with no background in cloud
computing. Given the availability of benchmarking tools, we are also investigating the integration of such tools to help on the data collection phase, while focusing on avoiding unnecessary executions to generate user insights.

\bibliographystyle{IEEEtran}

\balance
\bibliography{paper}

\begin{thebibliography}{10}
\providecommand{\url}[1]{#1}
\csname url@samestyle\endcsname
\providecommand{\newblock}{\relax}
\providecommand{\bibinfo}[2]{#2}
\providecommand{\BIBentrySTDinterwordspacing}{\spaceskip=0pt\relax}
\providecommand{\BIBentryALTinterwordstretchfactor}{4}
\providecommand{\BIBentryALTinterwordspacing}{\spaceskip=\fontdimen2\font plus
\BIBentryALTinterwordstretchfactor\fontdimen3\font minus
  \fontdimen4\font\relax}
\providecommand{\BIBforeignlanguage}[2]{{%
\expandafter\ifx\csname l@#1\endcsname\relax
\typeout{** WARNING: IEEEtran.bst: No hyphenation pattern has been}%
\typeout{** loaded for the language `#1'. Using the pattern for}%
\typeout{** the default language instead.}%
\else
\language=\csname l@#1\endcsname
\fi
#2}}
\providecommand{\BIBdecl}{\relax}
\BIBdecl

\bibitem{netto2018hpc}
M.~A. Netto, R.~N. Calheiros, E.~R. Rodrigues, R.~L. Cunha, and R.~Buyya,
  ``{HPC} cloud for scientific and business applications: taxonomy, vision, and
  research challenges,'' \emph{ACM Computing Surveys (CSUR)}, vol.~51, no.~1,
  pp. 1--29, 2018.

\bibitem{lamar2023evaluating}
K.~Lamar, A.~Goponenko, O.~Aaziz, B.~A. Allan, J.~M. Brandt, and D.~Dechev,
  ``Evaluating {HPC} job run time predictions using application input
  parameters,'' in \emph{Proceedings of the 17th ACM International Conference
  on Distributed and Event-based Systems (DEBS'23)}, 2023, pp. 127--138.

\bibitem{flores2021performance}
J.~Flores-Contreras, H.~A. Duran-Limon, A.~Chavoya, and S.~H. Almanza-Ruiz,
  ``Performance prediction of parallel applications: a systematic literature
  review,'' \emph{The Journal of Supercomputing}, vol.~77, pp. 4014--4055,
  2021.

\bibitem{wang2019novel}
Q.~Wang, J.~Li, S.~Wang, and G.~Wu, ``A novel two-step job runtime estimation
  method based on input parameters in {HPC} system,'' in \emph{2019 IEEE 4th
  International Conference on Cloud Computing and Big Data Analysis
  (ICCCBDA)}.\hskip 1em plus 0.5em minus 0.4em\relax IEEE, 2019, pp. 311--316.

\bibitem{tsafrir2007backfilling}
D.~Tsafrir, Y.~Etsion, and D.~G. Feitelson, ``Backfilling using
  system-generated predictions rather than user runtime estimates,'' \emph{IEEE
  Transactions on Parallel and Distributed Systems}, vol.~18, no.~6, pp.
  789--803, 2007.

\bibitem{yang2005cross}
L.~T. Yang, X.~Ma, and F.~Mueller, ``Cross-platform performance prediction of
  parallel applications using partial execution,'' in \emph{SC'05: Proceedings
  of the 2005 ACM/IEEE Conference on Supercomputing}.\hskip 1em plus 0.5em
  minus 0.4em\relax IEEE, 2005, pp. 40--40.

\bibitem{smith2007prediction}
W.~Smith, ``Prediction services for distributed computing,'' in \emph{2007 IEEE
  International Parallel and Distributed Processing Symposium}.\hskip 1em plus
  0.5em minus 0.4em\relax IEEE, 2007, pp. 1--10.

\bibitem{mariani2018predicting}
G.~Mariani, A.~Anghel, R.~Jongerius, and G.~Dittmann, ``Predicting cloud
  performance for {HPC} applications before deployment,'' \emph{Future
  Generation Computer Systems}, vol.~87, pp. 618--628, 2018.

\bibitem{betting2023oikonomos}
J.-H. Betting, D.~Liakopoulos, M.~Engelen, and C.~Strydis, ``Oikonomos: An
  opportunistic, deep-learning, resource-recommendation system for cloud
  {HPC},'' in \emph{2023 IEEE 34th International Conference on
  Application-specific Systems, Architectures and Processors (ASAP)}.\hskip 1em
  plus 0.5em minus 0.4em\relax IEEE, 2023, pp. 188--196.

\bibitem{lee2006user}
C.~B. Lee and A.~Snavely, ``On the user--scheduler dialogue: studies of
  user-provided runtime estimates and utility functions,'' \emph{The
  International Journal of High Performance Computing Applications}, vol.~20,
  no.~4, pp. 495--506, 2006.

\bibitem{bailey2005user}
C.~Bailey~Lee, Y.~Schwartzman, J.~Hardy, and A.~Snavely, ``Are user runtime
  estimates inherently inaccurate?'' in \emph{Job Scheduling Strategies for
  Parallel Processing: 10th International Workshop, JSSPP 2004, New York, NY,
  USA, June 13, 2004. Revised Selected Papers 10}.\hskip 1em plus 0.5em minus
  0.4em\relax Springer, 2005, pp. 253--263.

\bibitem{yang2023exploring}
W.~Yang, X.~Liao, D.~Dong, and J.~Yu, ``Exploring job running path to predict
  runtime on multiple production supercomputers,'' \emph{Journal of Parallel
  and Distributed Computing}, vol. 175, pp. 109--120, 2023.

\bibitem{brunetta2019selecting}
J.~R. Brunetta and E.~Borin, ``Selecting efficient cloud resources for {HPC}
  workloads,'' in \emph{Proceedings of the 12th IEEE/ACM International
  Conference on Utility and Cloud Computing}, 2019, pp. 155--164.

\bibitem{samuel2020a2cloud}
D.~Samuel, S.~Khan, C.~J. Balos, Z.~Abuelhaj, A.~D. Dutoi, C.~Kari, D.~Mueller,
  and V.~K. Pallipuram, ``A2cloud-rf: A random forest based statistical
  framework to guide resource selection for high-performance scientific
  computing on the cloud,'' \emph{Concurrency and Computation: Practice and
  Experience}, vol.~32, no.~24, p. e5942, 2020.

\bibitem{karakasis2020enabling}
V.~Karakasis, T.~Manitaras, V.~H. Rusu, R.~Sarmiento-P{\'e}rez, C.~Bignamini,
  M.~Kraushaar, A.~Jocksch, S.~Omlin, G.~Peretti-Pezzi, J.~P. Augusto
  \emph{et~al.}, ``Enabling continuous testing of hpc systems using reframe,''
  in \emph{Tools and Techniques for High Performance Computing: Selected
  Workshops, HUST, SE-HER and WIHPC, Held in Conjunction with SC 2019, Denver,
  CO, USA, November 17--18, 2019, Revised Selected Papers 6}.\hskip 1em plus
  0.5em minus 0.4em\relax Springer, 2020, pp. 49--68.

\bibitem{ramble2024}
\BIBentryALTinterwordspacing
{Ramble Development Team}, ``{Ramble: Reproducible And Measurable Benchmarks in
  a Layered Environment},'' 2024. [Online]. Available:
  \url{https://ramble.readthedocs.io/en/latest/}
\BIBentrySTDinterwordspacing

\bibitem{pavilion2}
\BIBentryALTinterwordspacing
{Pavilion 2 Development Team}, ``{Pavilion 2: A framework for running and
  managing tests on HPC systems},'' 2024. [Online]. Available:
  \url{https://pavilion2.readthedocs.io/en/latest/}
\BIBentrySTDinterwordspacing

\bibitem{gamblin2015spack}
T.~Gamblin, M.~LeGendre, M.~R. Collette, G.~L. Lee, A.~Moody, B.~R.
  De~Supinski, and S.~Futral, ``The spack package manager: bringing order to
  {HPC} software chaos,'' in \emph{Proceedings of the International Conference
  for High Performance Computing, Networking, Storage and Analysis}, 2015, pp.
  1--12.

\bibitem{hoste2012easybuild}
K.~Hoste, J.~Timmerman, A.~Georges, and S.~De~Weirdt, ``Easybuild: Building
  software with ease,'' in \emph{2012 SC Companion: High Performance Computing,
  Networking Storage and Analysis}.\hskip 1em plus 0.5em minus 0.4em\relax
  IEEE, 2012, pp. 572--582.

\bibitem{droge2023eessi}
B.~Dr{\"o}ge, V.~Holanda~Rusu, K.~Hoste, C.~van Leeuwen, A.~O'Cais, and
  T.~R{\"o}blitz, ``Eessi: A cross-platform ready-to-use optimised scientific
  software stack,'' \emph{Software: Practice and Experience}, vol.~53, no.~1,
  pp. 176--210, 2023.

\end{thebibliography}

\end{document}